\documentclass[preprint,tightenlines,aps,nofootinbib,showpacs]{revtex4}

\usepackage{epsf}
\usepackage{epsfig}
\usepackage{graphicx}
\usepackage[dvips]{color}

\newcommand{\notE}{ \hbox{{$E$}\kern-.60em\hbox{/}}}
\newcommand{\notp}{\ \hbox{{$p$}\kern-.43em\hbox{/}}}
\def\D0{\mbox{D\O}}

\newcommand{\eps}{\epsilon}

\includegraphics{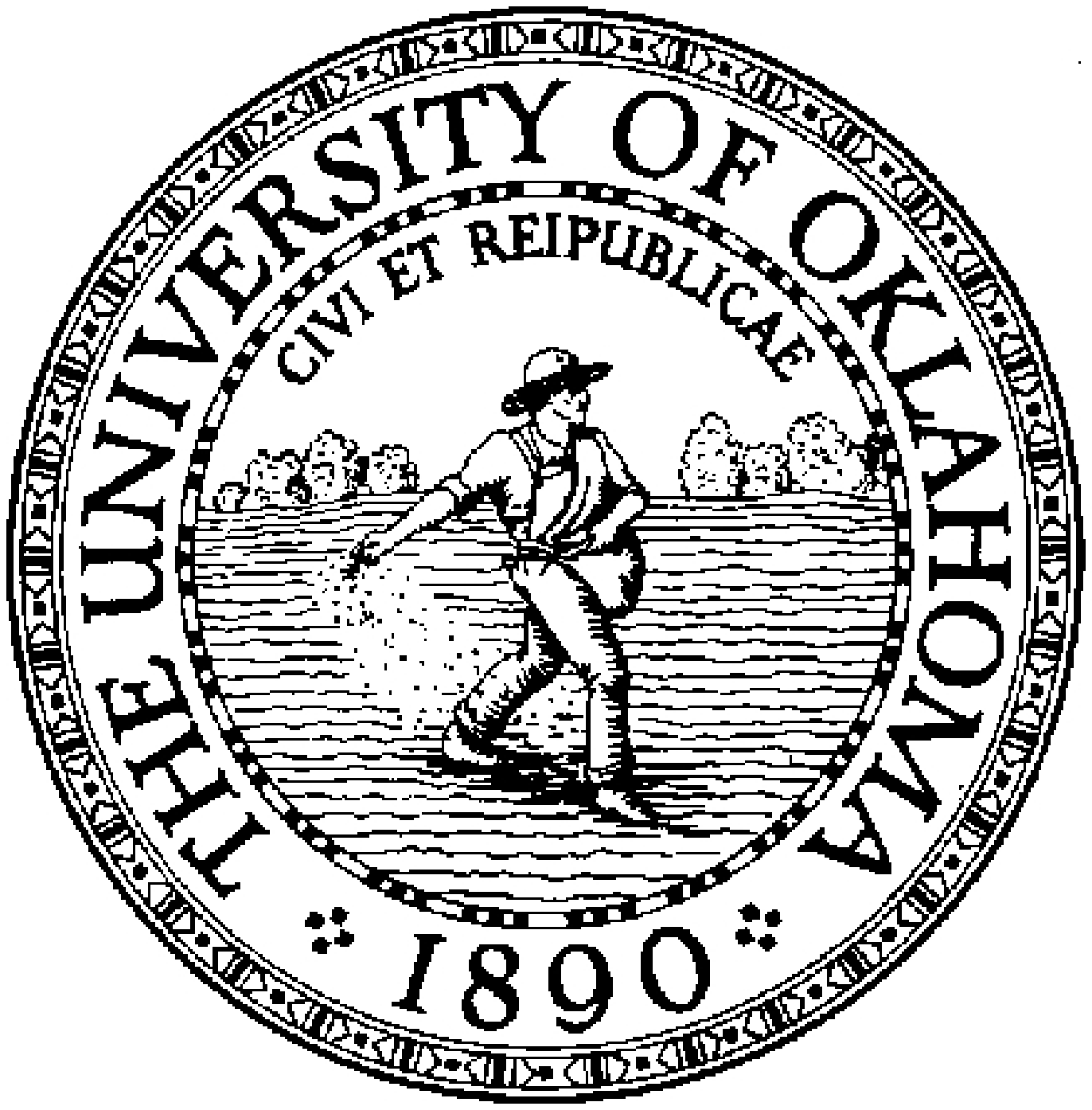}

\preprint{\font\fortssbx=cmssbx10 scaled \magstep2
\hbox to \hsize{
\hskip1.2in 
\hbox{\fortssbx The University of Oklahoma}
\hskip0.2in $\vcenter{
                      \hbox{\bf arXiv: [hep-ph]}
                      \hbox{\bf OU-HEP-181222}
                      \hbox{December 2018}}$ }
}

\begin{document}


\title{\vspace*{0.7in}
Charming Top Decays with Flavor Changing Neutral Higgs Boson 
and WW at Hadron Colliders}

\author{
Rishabh Jain\footnote{E-mail address: Rishabh.Jain@ou.edu} and 
Chung Kao\footnote{E-mail address: Chung.Kao@ou.edu}
}

\affiliation{
Homer L. Dodge Department of Physics and Astronomy,
University of Oklahoma, Norman, OK 73019, USA }

\date{\today}

\bigskip

\begin{abstract}

We investigate the prospects for discovering a top quark decaying 
into one light Higgs boson ($h^0$) along with a charm quark ($c$) 
in top quark pair production at the CERN Large Hadron Collider (LHC) 
and future hadron colliers. 
A general two Higgs doublet model is adopted to study the signature 
of flavor changing neutral Higgs (FCNH) interactions with $t \to c h^0$, 
followed by $h^0 \to WW^* \to \ell^+ \ell^- +\notE_T$, 
where $h^0$ is the CP-even Higgs boson and $\notE_T$ stands for 
missing transverse energy from neutrinos.
We study the discovery potential for this FCNH signal and physics
background from dominant processes with realistic acceptance cuts 
as well as tagging and mistagging efficiencies. 
Promising results are found for the LHC running
at 13 TeV and 14 TeV center-of-mass energy as well as future pp colliders 
at 27 TeV and 100 TeV.

\end{abstract}

%


\maketitle

\newpage

\section{Introduction}

The Standard Model has been very successful in explaining 
almost all experimental data to date, culminating in the discovery of 
the the long awaited standard Higgs boson at the CERN 
Large Hadron Collider (LHC)~\cite{Aad:2012tfa,Chatrchyan:2012xdj}. 
The most important experimental goals of the LHC, future high energy 
hadron colliders, and the International Linear Collider (ILC) 
are to study the Higgs properties and to search for new physics 
beyond the Standard Model (SM) 
including additional Higgs bosons and flavor changing neutral Higgs 
(FCNH) interactions.

In the Standard Model there is one Higgs doublet, which generates 
masses for both vector bosons and fermions. There is no explanation for 
the large differences among Yukawa couplings of fermions with the 
Higgs boson.
In addition, there are no flavor changing neutral currents (FCNC) mediated by
gauge interactions or by Higgs interactions at the tree level.
The top quark is the most massive elementary particle ever
discovered. The fact that the Higgs boson ($h^0$) is lighter than 
the top quark ($m_t > M_h$) makes it possible for the top quark 
to decay into the Higgs boson along with a charm quark ($t \to c h^0$) 
kinematically. 
At the one loop level, the branching fraction of $t \to c h^0$ is
$3 \times 10^{-15}$ for $M_h = 125$ GeV \cite{
AguilarSaavedra:2004wm,Mele:1998ag,Eilam:1990zc}.
If this decay mode is detected in the near future, it would indicate 
a large tree-level coupling or a significant enhancement from beyond SM 
loop effects.

A general two Higgs doublet model (2HDM) usually contains flavor changing
neutral Higgs (FCNH) interactions if there is no discrete symmetry
to turn off tree-level FCNC~\cite{Glashow:1976nt,Guide}.
In 1991, it was pointed out that top-charm FCNH coupling could be 
prominent~\cite{Hou:1991un} if the Yukawa couplings of fermions 
and the Higgs boson are comparable to the geometric mean of 
their mass~\cite{Cheng:1987rs}.
A special two Higgs doublet model for the top quark 
(T2HDM)~\cite{Das:1995df} might provide a reasonable explanation 
why the top quark is much more massive than other elementary fermions. 
In the T2HDM, top quark is the only elementary fermion acquiring
its mass from a special Higgs doublet ($\phi_2$) with
a large vacuum expectation value ($v_2 \gg v_1$).
Since the up and charm quarks couple to another Higgs doublet ($\phi_1$),
there are FCNH interactions among the up-type quarks.
The down type quarks have the same interactions as those in the SM.

In a general two Higgs doublet model, there are five physical Higgs bosons: 
two CP-even scalars $h^0$ (lighter) and $H^0$ (heavier), 
a CP-odd pseudoscalar ($A^0$), and 
a pair of singly charged Higgs boson ($H^\pm$). 
To study FCNH interactions in a general 2HDM, 
we employ the following Lagrangian 
with Higgs bosons and fermions~\cite{Davidson:2005cw,Mahmoudi:2009zx},
\begin{eqnarray}
{\cal L}_Y &=& \frac{-1}{\sqrt{2}} \sum_{\scalebox{0.6}{F=U,D,L}}
 \bar{F}\left\{  \left[ \kappa^Fs_{\beta-\alpha}+\rho^F c_{\beta-\alpha} \right] h^0 +
 \left[ \kappa^Fc_{\beta-\alpha}-\rho^Fs_{\beta-\alpha} \right] H^0 - i \, {\rm sgn}(Q_F)\rho^F A^0 \right\} P_R F \nonumber \\
 && -\bar{U} \left[ V \rho^D P_R - \rho^{U\dagger} V P_L \right] D H^+
  -\bar{\nu} \left[ \rho^L P_R \right] L H^+ + {\rm H.c.} \, 
\end{eqnarray}
where $P_{L,R} \equiv ( 1\mp \gamma_5 )/2$,
$c_{\beta-\alpha} = \cos(\beta-\alpha)$,
$s_{\beta-\alpha} = \sin(\beta-\alpha)$,
$\alpha$ is the mixing angle between neutral Higgs scalars, 
$\tan\beta \equiv v_2/v_1$~\cite{Guide}, 
$Q_F$ is the charge of a fermion,and 
$\kappa$~matrices are diagonal and fixed by
fermion masses to $\kappa^F = \sqrt{2}m_F/v$ with $v \simeq 246$~GeV, while
$\rho$ matrices have both diagonal and off-diagonal
elements with free parameters.

The LHC has become a top quark factory. The production cross section of 
top quark pair ($\sigma_{tt}$) is approximately 820 pb in pp collisions 
with a 13 TeV center-of-mass energy (CM) energy ($\sqrt{s}$), and it becomes 
$\sigma_{tt} \simeq 970$ pb at $\sqrt{s} = 14$ TeV~\cite{
Nason:1987xz,Kidonakis:2010dk,Ahrens:2011px,Cacciari:2011hy,Czakon:2013goa}. 
For an integrated luminosity of ${\cal L} = 100$ fb$^{-1}$ 
at $\sqrt{s} = 13$ TeV, the LHC has produced more than $8 \times 10^7$ 
top quark pairs ($t\bar{t}$)
for $m_t \simeq 173.2$ GeV~\cite{Aaboud:2016pbd,Sirunyan:2017uhy}.
For the same integrated luminosity at $\sqrt{s} = 14$ TeV,
the number of ($t\bar{t}$) pairs generated would increase to about 
$1 \times 10^8$.
Thus, the LHC will provide great opportunities to study electroweak symmetry 
breaking as well as other important properties of the top quark 
and the Higgs boson.

Most ATLAS and CMS measurements of the 125 GeV Higgs boson ($h^0$) 
are consistent with expectations for the Standard Model. 
The branching fractions of the standard Higgs boson are presented in
Table I~\cite{CERNHWG,Aad:2015gba,TheATLASandCMSCollaborations:2015bln}. 
In a general two Higgs doublet model, 
let us consider the light Higgs scalar ($h^0$) as the SM Higgs boson 
in the alignment limit~\cite{Gunion:2002zf,Carena:2013ooa}.
 
\begin{table}[htb]
\label{HiggsBF}
\caption{Branching fractions and partial decay widths of the light 
CP-even Higgs boson ($h^0$) of a general two Higgs 
doublet model in the alignment limit ($h^0 \simeq h^0_{\rm SM}$). 
For simplicity, let us take $\rho_{ff} \simeq \kappa_f = \sqrt{2}m_f/v$.  
Widths are in MeV units, with 
$\Gamma_{h^0}^{\rm SM} \simeq 4.1$ MeV~\cite{CERNHWG}. }
\begin{tabular}{cllcc}
\hline\hline
Decay Channel $\quad$ & ${\cal B}^{\rm SM} \quad$ & $\Gamma$ [MeV] & Comment \\
\hline
$bb$       & 57.5\%    & \ 2.35 & $\rho_{bb} \simeq \kappa_b$ \\
$WW^*$     & 21.6\%    & \ 0.89 & $\sin(\beta - \alpha) \simeq 1$ \\
$gg$       & $\;\, 8.56\%$ & \ 0.35 & $\rho_{tt} \simeq \kappa_t \sim 1$ \\
$\tau\tau$ & $\;\, 6.30\%$ & \ 0.26 & $\rho_{\tau \tau} \simeq \kappa_{\tau} $ \\
$ZZ^*$     & $\;\, 2.67\%$ & \ 0.11 & $\sin(\beta - \alpha) \simeq 1$ \\
$\gamma\gamma$ & $\;\, 0.23\%$ & \ 0.094  & $W$-loop and fermion loops. \\
\hline\hline
\end{tabular}
\end{table}

It is clear that the most probable decay channels are 
$b\bar{b}$ and $WW$ with branching fractions 
${\cal B}(h^0 \to b\bar{b}) \simeq 0.58$ and 
${\cal B}(h^0 \to WW^*) \simeq 0.22$ as show in Table I.
However, the light Higgs boson was first discovered with 
$h^0 \to \gamma\gamma$ 
and $h^0 \to ZZ^* \to 4\ell$, because these channels have 
less background and better mass resolutions.
In the past few years, several theoretical studies and experimental 
searches have been completed for the charming top FCNH decay $t \to ch^0$ with 
(a) $h^0 \to b\bar{b}$~\cite{
Kao:2011aa,Atwood:2013ica,Sirunyan:2017uae,Banerjee:2018fsx}, 
(b) $h^0 \to ZZ^*$~\cite{Chen:2013qta}, 
(c) $h^0 \to \gamma\gamma$~\cite{Aaboud:2017mfd,Banerjee:2018fsx}, and 
(d) Higgs decays into multileptons~\cite{
Craig:2012vj,CMS:2014qxa,Khachatryan:2014jya}.
Recently, the ATLAS collaboration has placed tight limits on the FCNH 
branching fraction for $t \to ch^0$ and the Yukawa coupling $\lambda_{tch}$ 
with Higgs boson decaying into multileptons~\cite{Aaboud:2018pob}
\begin{equation}
{\cal B}(t \to ch^0) \le 0.16\% \, , 
\quad {\rm and} \quad \lambda_{tch} \le 0.077 \, ,
\end{equation}
for the effective Lagrangian
\begin{equation}
{\cal L}_{\rm eff}
 = -\frac{\lambda_{tch}}{\sqrt{2}} \bar{c}t h^0 +{\rm H.c.} \, .
\end{equation}
The LHC limits for the branching ratios can be translated to a limit
on the flavor changing Yukawa coupling by a simple rescaling.
It is a good approximation to consider a simple numerical relation between 
the FCNH Yukawa coupling ($\lambda_{tch}$) and the branching fraction of 
$t \to c h^0$~\cite{ATLAS2013tch}
\begin{equation}
\lambda_{tch} \simeq 1.92 \times \sqrt{{\cal B}(t \to c h^0)} \, .
\end{equation}

In this article, we focus on the discovery potential of the LHC in the search
for the FCNH top decay $t \to c h^0$ followed by 
$h^0 \to W W^* \to \ell^+ \ell^- \nu \bar{\nu}$. 
We have evaluated production rates with full tree-level matrix elements
including Breit-Wigner resonances for both the signal and the physics
background. In addition, we optimize the acceptance cuts to effectively
reduce the background with realistic $b$-tagging and mistagging
efficiencies. 
Promising results are presented for the LHC with $\sqrt{s} =13$ TeV 
and $\sqrt{s} = 14$ TeV as well as 
for future hadron colliders at $\sqrt{s} = 27$ TeV and 100 TeV, for 
High Luminosities (HL)~\cite{
Barletta:2013ooa,Tomas:2016,Zimmermann:2017bbr,Shiltsev:2017tjx}  
of $L = 300$ fb$^{-1}$ and 3000 fb$^{-1}$.
Section II shows the production cross sections for the Higgs signal
and the dominant background, as well as our strategy to determine
the reconstructed masses for the top quark and the Higgs boson.
Realistic acceptance cuts are discussed in Section III.
Section IV presents the discovery potential at the LHC
for $\sqrt{s} = 13$ TeV and 14 TeV, 
as well as for future hadron colliders with
for $\sqrt{s} = 27$ TeV and 100 TeV.
Our optimistic conclusions are drawn in Section V.

\section{The Higgs Signal and Physics Background}

In this section we present the cross section for the FCNH Higgs signal 
in pp collisions 
($pp \to t\bar{t} \to tch^0 \to bjj \, c\ell\ell\nu\bar{\nu} +X, \ell = e, \mu$) 
as well as for the dominant physics background processes.
Figure 1 shows the Feynman diagram of top quark pair production in pp 
collisions from gluon fusion and quark-antiquark fusion, 
followed by one top quark decaying into a Higgs boson 
and a charm quark,  
while the other top quark decays into $bW \to bjj$.


\begin{figure}[htb]

\begin{center}
\includegraphics[width=72mm]{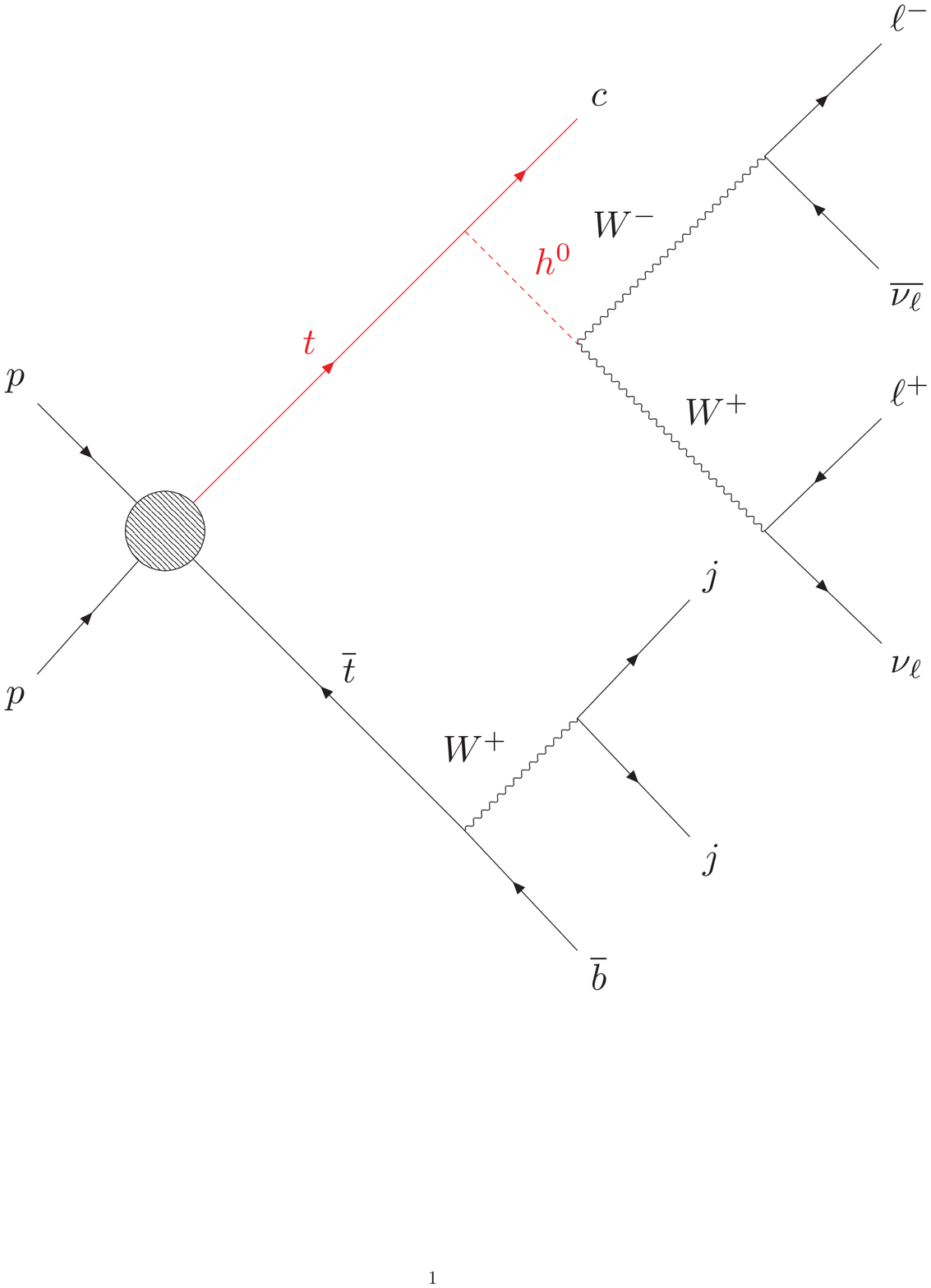}
\caption{Feynman diagram for  
$pp \to t\bar{t} \to bjj \, ch^0 + X
    \to bjj  \, c\ell^+ \ell^- \nu\bar{\nu} X$, 
where $\ell = e$ or $\mu$.}
\end{center}

\end{figure}

\subsection{The Higgs Signal in Top Decay}

Applying the Lagrangian in Eq.~[1] with general Yukawa interactions 
for the light Higgs boson and fermions, 
we obtain the decay width of $t \to ch^0$ 
\begin{equation}
 \Gamma_{t \to c h^0}
 = \frac{c_{\beta-\alpha}^2 m_t}{32 \pi}
   \left[ (1 + r_c^2 - r_h^2)\frac{(|\rho_{ct}|^2 + |\rho_{tc}|^2)}{2} + r_c(\rho_{tc}^*\rho_{ct}^* + \rho_{tc}\rho_{ct)} \right] 
   \lambda^{1/2}(1,r_c^2,r_h^2)
\end{equation}
where $c_{\beta-\alpha} = \cos(\beta-\alpha)$, $r_h = M_h/m_t$, 
$r_c = m_c/m_t$, and 
\begin{equation}
\lambda(x,y,z) = x^2 +y^2 +z^2 -2xy -2xz -2yz \, .
\end{equation}

Let us define two variables, 
\begin{equation}
 \tilde{\rho}_{tc} = \sqrt{\frac{|\rho_{tc}|^2 + |\rho_{ct}|^2}{2}}, 
\quad {\rm and} \quad
 \epsilon_c = \rho_{tc}^*\rho_{ct}^* + \rho_{tc}\rho_{ct} \, ,
\end{equation}
Combining LHC Higgs data and $B$ physics, a recent study found constraints 
$\rho_{tc} \le 1.5$ and $\rho_{ct} \le 0.1$~\cite{Altunkaynak:2015twa}. 
That implies $\eps_c \alt 0.2 \tilde{\rho}_{tc}$ for $\rho_{tc} \simeq 1$.
Hence we can write our decay width as,
\begin{equation}
 \Gamma_{t \to c h^0} = \frac{c_{\beta-\alpha}^2 m_t}{32 \pi}
     [ (1 +r_c^2 -r_h^2)|\tilde{\rho}_{tc}|^2 +\epsilon_c r_c ]
     \times \lambda^{1/2}(1,r_c^2,r_h^2) \, .
\end{equation}
Using $m_t = 173.2$ GeV, $M_h = 125.1$ GeV and 
$m_c = 1.42$ GeV~\cite{CERNHWG}, we obtain 
\begin{equation}
 \Gamma_{t \to c h^0} = \frac{c_{\beta-\alpha}^2 m_t}{32 \pi}
     [ 0.48 |\tilde{\rho}_{tc}|^2 + 0.008 \epsilon_c) ]
     \times \lambda^{1/2}(1,r_c^2,r_h^2) \, .
\end{equation}

Since we have $m_c \ll m_t$, $r_c \ll 1$, 
and $\epsilon_c \alt 0.2 |\tilde{\rho}_{tc}|$, 
it is a very good approximation to consider 
\begin{equation}
 \Gamma_{t \to c h^0} \simeq  \frac{c_{\beta-\alpha}^2 m_t}{32 \pi}
     [(1 - r_h^2)|\tilde{\rho}_{tc}|^2 ]
     \times \lambda^{1/2}(1,r_c^2,r_h^2) \, .
\end{equation}
For typical values of parameters $\cos(\beta-\alpha) = 0.1$, 
$|\rho_{tc}| \sim 1$ and $|\rho_{ct}| \sim 0.1$, we have 
\begin{equation}
 \Gamma_{t \to c h^0}
  \simeq 0.394 (c_{\beta-\alpha}^2|\tilde{\rho}_{tc}|^2)
  \simeq 0.00197 \; {\rm GeV} \, ,
\end{equation}
and 
\begin{equation}
    {\cal B}{(t \to c h^0)} \simeq 0.00132 \, .
\end{equation}

For simplicity, we may adopt the following effective Lagrangian to study 
FCNH Yukawa interactions for the light CP-even Higgs boson ($h^0$) 
with the top quark ($t$) and the charm quark ($c$)  
\begin{eqnarray}
{\cal L}
 = -g_{htc} \bar{c}t h^0 +{\rm H.c.},
\end{eqnarray}
where 
\begin{eqnarray}
g_{htc} = \frac{1}{\sqrt{2}}\tilde{\rho}_{tc} \cos(\beta-\alpha) 
        = \frac{1}{\sqrt{2}}\lambda_{tch} \, .
\end{eqnarray}
It is the effective coupling of the FCNH Yukawa coupling.

Then the decay width for $t \to c h^0$ \cite{Hou:1991un} becomes 
\begin{eqnarray}
\Gamma(t \to c\phi^0)
 =  \frac{|g_{htc}|^2}{16\pi}\times (m_t)\times
      [ 1 + r_c^2 -r_h^2 ]
      \times \sqrt{1-(r_h+r_c)^2}\sqrt{1-(r_h-r_c)^2} \, .
\end{eqnarray}

We assume that the total decay width of the top quark is
\begin{eqnarray}
\Gamma_t = \Gamma( t\to bW ) +\Gamma( t \to c h^0 ) \, .
\end{eqnarray}
Then the branching fraction of $t \to c h^0$ becomes
\begin{eqnarray}
{\cal B}(t \to c h^0) = \frac{ \Gamma(t\to c h^0) }{ \Gamma_t } \, .
\end{eqnarray}

As a case study, let us take the FCNH Yukawa couplings
to be the geometric mean of the Yukawa couplings of the quarks 
that is also known as the Cheng-Sher (CS) Ansatz~\cite{Cheng:1987rs} 
\begin{equation}
g_{htc}({\rm CS}) = \frac{ \sqrt{m_t m_c} }{v} \simeq 0.0637,
\end{equation}
or
\begin{equation}
\lambda_{tch}({\rm CS}) = \sqrt{2} \, g_{htc}({\rm CS})
              = \frac{\sqrt{2 m_t m_c}}{v} \simeq 0.0901 \, ,
\end{equation}
with $m_t = 173.2$ GeV and $m_c = 1.42$ GeV.
Then the branching fraction of $t \to c h^0$ becomes
${\cal B}(t\to c h^0) = 2.2 \times 10^{-3}$ for $M_{h} = 125.1$ GeV. 
In general, we will consider 
$g_{htc} = \tilde{\rho}_{tc}\cos(\beta-\alpha)/\sqrt{2}$ with 
$\tilde{\rho}_{tc}$ and $\cos(\beta - \alpha)$ as free parameters.

We employ the programs MadGraph~\cite{Alwall:2011uj,Hagiwara:2008jb} and
HELAS~\cite{HELAS} to evaluate the exact matrix element
for the FCNH signal in top decays from gluon fusion
and quark-antiquark annihilation,
\begin{eqnarray}
gg, q\bar{q} & \to & t \bar{t}
 \to t \bar{c} h^0 \to bjj \bar{c} \ell^+ \ell^- \nu\bar{\nu} \, , 
 \quad {\rm and} \, , 
 \nonumber \\
gg, q\bar{q} & \to & \bar{t} t
 \to \bar{t} c h^0 \to \bar{b}jj c \ell^+ \ell^- \nu\bar{\nu} \, ,
\end{eqnarray}
where $\ell = e$ or $\mu$.
The cross section of the Higgs signal in FCNH top decays at the LHC 
and future hadron colliders for
$pp \to t\bar{t}
    \to t c h^0 \to bjj \, c\ell^+ \ell^- \nu \bar{\nu} +X$
is evaluated with the parton distribution functions of 
CT14LO~\cite{Dulat:2015mca,Gao:2013xoa} with a common value 
$Q = M_{t\bar{t}} =$ the invariant mass of $t\bar{t}$, 
for the renormalization scale ($\mu_R$) and the factorization scale ($\mu_F$). 
This choice of scale leads to a K factor of approximately 1.8 
for top quark pair production. We have used the computer program 
Top++~\cite{Czakon:2013goa} to evaluate higher order corrections. 
In addition, we have checked the tree-level signal cross section 
with narrow width approximation. That is, the cross section
$\sigma(pp \to t\bar{t} \to t c h^0 \to bjj \, c\ell^+ \ell^- \nu\bar{\nu} +X)$
is calculated as the product of cross section times branching
fractions:
\begin{equation}
\sigma(pp \to t\bar{t} \to bjj \bar{t} + X )
 \times {\cal B}(t\to c h^0) \times {\cal B}(h^0 \to W^+ W^-)
 \times \left[ {\cal B}({W \to \ell \nu_{\ell}}) \right]^2 \, .
\end{equation}

In our analysis, we consider the FCNH signal from both
$t\bar{t} \to t\bar{c} h^0
          \to bjj \, \bar{c} \ell^+ \ell^- \nu_{\ell} \bar{\nu_{\ell}}$
and
$t\bar{t} \to c h^0 \bar{t}
          \to \bar{b}jj \, c \ell^+ \ell^- \nu_{\ell} \bar{\nu_{\ell}}$.
In every event, we require that there should be one $b$ jet and 
three light jets ($j = u, d, s, c$, or $g$ in physics background).
In addition, there are two leptons ($\ell = e$ or $\mu$) and neutrinos, 
which will  be lead to missing transverse energy ($\notE_T$).
Unless explicitly specified,
$q$ generally denotes a quark ($q$) or an anti-quark ($\bar{q}$)
and $\ell$ will represent a lepton ($\ell^-$) or anti-lepton ($\ell^+$).
That means our FCNH signal leads to the final state of
$bjj \, c\ell^+\ell^- \nu_{\ell} \bar{\nu_{\ell}}$ or 
$bjjj\ell^+\ell^- +\notE_T$.

\subsection{The Physics Background}

The dominant physics background to the final state of 
$bjj c \ell^+ \ell^- \nu \bar{\nu}$ comes from top quark pair production 
along with two light jets ($t\bar{t}jj$), 
$pp \to t\bar{t} jj \to b\bar{b} jj W W
    \to b\bar{b} jj \ell^+\ell^- \nu\bar{\nu} +X$, 
where every top quark decays into a $b-$quark as well as 
a $W$ boson ($W \to \ell\nu$) 
and a $b$-jet is mis-identified as a $c$-jet. 
We have also considered backgrounds from
$pp \to t\bar{t} W \to b\bar{b} jj W W 
    \to b\bar{b} jj \ell^+\ell^- \nu\bar{\nu} +X$
with one $W$ boson decaying into $jj$, and
$pp \to b \bar{b} jj W W 
    \to b \bar{b} jj  \ell^+ \ell^- \nu\bar{\nu} + X$, 
excluding the contribution from $t \bar{t} j j$ and $t\bar{t} W$. 
In addition, we have included 
$pp \to c \bar{c} jj W W \to c \bar{c} jj \ell^+ \ell^- \nu\bar{\nu} + X$ and
$pp \to jjjj W W \to jjjj \ell^+ \ell^- \nu\bar{\nu} + X$ 
where $j = u, d, s$, or $g$. 
We evaluate the cross section of physics background in pp collisions with 
proper tagging and mistagging efficiencies.
In our analysis, we adopt updated ATLAS tagging 
efficiencies~\cite{ATLAS:2018bpl,Scodellaro:2017wli}: 
the $b$ tagging efficiency is$\sim 70 \% $,
the probability that a $c$-jet is mistagged as a $b$-jet ($\epsilon_c$)
is approximately $14\%$, while
the probability that any other jet is mistagged as a $b$-jet ($\epsilon_j$)
is $1\%$.

\subsection{Mass Reconstruction}

In this subsection, we demonstrate that the proposed Higgs signal 
comes from top quark pair production with 
$t\bar{t} \to bjj \, c h^0 \to bjj \, c\ell^+\ell^- +\notE_T$.
We discuss our strategy to determine the reconstructed 
top mass as the invariant mass of $bjj$ from $t \to bW \to bjj$ 
along with another top quark decays into a Higgs boson and 
a charm quark $t \to c h^0$.
Furthermore, we employ cluster transverse mass distributions 
for $\ell^+\ell^-$ and $c\ell^+\ell^-$ 
with missing transverse energy ($\notE_T$) from neutrinos.  
These distributions have broad peaks near $M_h$ and $m_t$ respectively 
as the kinematic characteristics of 
$t \to c h^0 \to c \ell^+\ell^- +\notE_T$. 
Applying suitable cuts on the cluster transverse mass $M_T(\ell\ell,\notE_T)$ 
as well as $M_T(c\ell\ell,\notE_T)$, we can greatly reduce 
the physics background and enhance the statistical significance 
for the Higgs signal.


\begin{figure}[htb]
\begin{center}
\includegraphics[width=80mm]{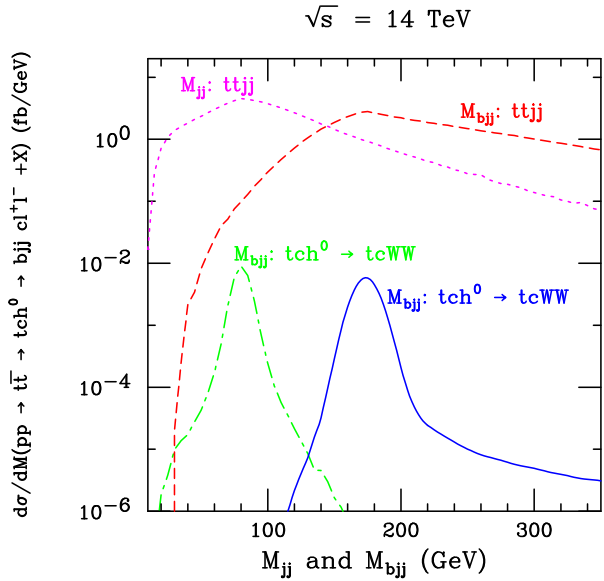}
\caption{
Invariant mass distributions ($d\sigma/dM$) of $j_1 j_2$ (green dotdash),
and  $b j_1 j_2$ (blue solid), for the Higgs signal in pp collisions, 
$d\sigma/dM(pp \to t\bar{t} \to t ch^0
  \to t c W W \to b j j c \ell^+ \ell^- +\notE_T +X$ (fb/GeV), 
with basic cuts defined in Eq.~[22].
Also shown are the invariant mass distributions $d\sigma/dM_{j1 j2}$ 
(magenta dot) and $d\sigma/dM_{b j1 j2}$ (red dash) 
for the dominant physics background from $t\bar{t}jj$.  
}
\label{fig:invariant_mass} 
\end{center}
\end{figure}

In our analysis, we assume that the FCNH signal comes from top quark
pair production with one top quark decaying into a charm quark and 
a Higgs boson ($t \to ch^0 \to cWW \to c \ell^+ \nu \ell^- \bar{\nu}$) 
while the other decays hadronically ($t \to bW \to b j j$).
In every event, there is one tagged $b$-jet and three light jets.
Let us choose the pair of light jets that minimize 
$|M_{jj}-m_W|$ and $|M_{bjj}-m_t|$
as $j_1 j_2$ and label the other jet as $j_3 \simeq c$.
That means, for a correctly reconstructed event, $j_1$ and $j_2$ 
are the products of a $W$ decay such that their invariant mass distribution 
peaks at $M_{j_1 j_2} \simeq m_W$.
For a background event, one $b$ is likely coming from the top decay
$t \to bW \to b j j$ while the other is either a mistagged $c$ or
a light quark jet coming from $W$ decay, or a real $b$ quark coming
from the decay of $\bar{t}$.

We present the invariant mass distributions for $M_{j_1 j_2}$ and 
$M_{b j_1 j_2}$ in FIG.~2 for the Higgs signal ($t\bar{t} \to tch^0$) and 
the dominant background ($t\bar{t}jj$) with basic Cuts 
from CMS \cite{CMS:2016rnk}:
\begin{eqnarray}
& & {\rm (a)} \; p_T (b,j) > 25 \; {\rm GeV} \, , \nonumber \\
& & {\rm (b)} \; p_T(\ell_1) > 25 \; {\rm GeV} \, ,
                 p_T(\ell_2) > 15 \; {\rm GeV} \, , \nonumber \\
& & {\rm (c)} \; \notE_T > 25 \; {\rm GeV} \, , \nonumber \\
& & {\rm (d)} \; |\eta|(j,\ell)| < 2.4 \, , \quad {\rm and} \nonumber \\
& & {\rm (e)} \; |\Delta{R}(jj,\ell\ell,j\ell)| > 0.4 \, , 
\label{basic_cuts}
\end{eqnarray}
where $p_T(\ell_1) \ge p_T(\ell_2)$ and 
$\Delta R \equiv \sqrt{ (\Delta\phi)^2+(\Delta\eta)^2 }$.
It is clear to see that $M_{j_1 j_2}$ distribution peaks at $m_W$ 
while $d\sigma/dM_{bjj}$ has a peak at $m_t$.

In a good reconstruction, the remaining light jet, $j_3 \sim c$ should
reproduce the top quark mass with the momenta of charged leptons and neutrinos.
To reconstruct the Higgs mass and top mass for 
$t \to c h^0 \to c \ell^+\ell^- +\notE_T$, 
we use cluster transverse mass $M_T(\ell\ell,\notE_T)$ and 
$M_T(c\ell\ell,\notE_T)$~\cite{Colphy:1987,Han:1998ma}, defined below,
\begin{eqnarray}
M_T^2(\ell\ell,\notE_T)
  =  \left( \sqrt{ p_T^2(\ell\ell) +M_{\ell\ell}^2 } + \notE_T \right)^2
      -( \vec{p}_T(\ell\ell) +\vec{\notE}_T )^2  \, , 
\end{eqnarray}
and
\begin{eqnarray}
M_T^2(c\ell\ell,\notE_T)
  =  \left( \sqrt{ p_T^2(c\ell\ell) +M_{c\ell\ell}^2 } + \notE_T \right)^2
      -( \vec{p}_T(c\ell\ell) + \vec{\notE}_T )^2  \, ,
\end{eqnarray}
where $p_T(\ell\ell)$ or $p_T(c\ell\ell)$ is the total transverse momentum 
of all the visible particles and $M_{\ell\ell}$ or $M_{c\ell\ell}$ 
is the invariant mass.


\begin{figure}[htb]
\begin{center}
\includegraphics[width=80mm]{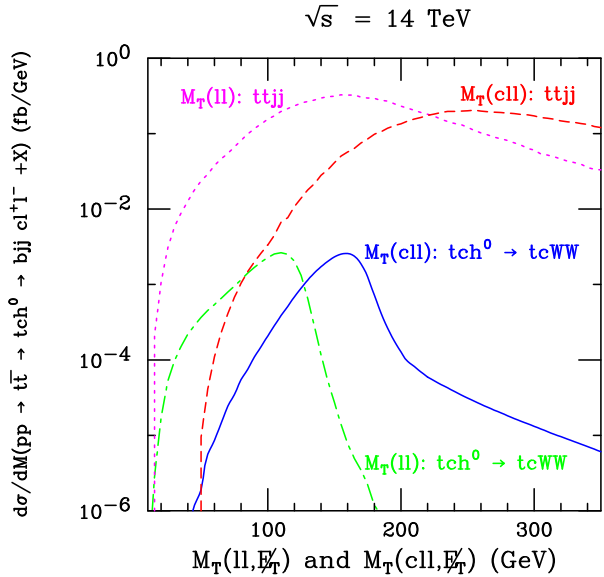}
\caption{
Cluster transverse mass distributions ($d\sigma/dM_T$) of 
$\ell^+\ell^-$ (green dotdash) and $c\ell^+\ell^-$ (blue solid) 
for the Higgs signal in pp collisions, 
$d\sigma/dM_T(pp \to t\bar{t} \to t ch^0
  \to t c W W \to b j j c \ell^+ \ell^- +\notE_T +X$ (fb/GeV), 
with basic cuts defined in Eq.~[22], as well as 
$|M_{jj} - m_W| \leq 0.15 \times m_W$ and 
$|M_{bjj} - m_t| \leq 0.20 \times m_t$.
Also shown are the cluster transverse mass distributions 
$d\sigma/dM_T(\ell\ell,\notE_T)$ (magenta dot) and 
$d\sigma/dM_T(c\ell\ell,\notE_T)$ (red dash)
for the dominant physics background from $t\bar{t}jj$.  
}
\label{fig:transverse_mass} 
\end{center}
\end{figure}

Figure~3 presents the cluster transverse mass distributions 
($d\sigma/dM_T(\ell\ell,\notE_T$) and ($d\sigma/dM_T(c\ell\ell,\notE_T$) 
for the Higgs signal in pp collisions, 
$d\sigma/dM_T(pp \to t\bar{t} \to t ch^0
  \to t c W W \to b j j c \ell^+ \ell^- +\notE_T +X$ (fb/GeV), 
with basic cuts defined in Eq.~[22], as well as 
$|M_{jj} - m_W| \leq 0.15 \times m_W$ and 
$|M_{bjj} - m_t| \leq 0.20 \times m_t$.
the cluster transverse mass distributions 
for $\ell^+\ell^-$ and 
$c\ell^+\ell^-$ for the Higgs signal ($t\bar{t} \to tch^0$) and 
the dominant background ($t\bar{t}jj$) with basic cuts defined in Eq.~[22] 
as well as invariant mass cuts
Note that $d\sigma/dM_T(\ell\ell,\notE_T$ peaks near $M_h$ while 
$d\sigma/dM_T(c\ell\ell,\notE_T$ has a peak near $m_t$.
 
It is clear that there are pronounced peaks at $m_W$ and $m_t$ 
in the invariant mass distributions of jets as shown in FIG.~2. 
We can also see broad peaks near $M_h$ and $m_t$ 
in the cluster transverse mass distributions:
\begin{eqnarray}
M^*_{j_1 j_2} & \simeq & m_W \, , \nonumber \\
M^*_{b j_1 j_2} & \simeq & m_t \, , \nonumber \\
M^*_T(\ell\ell,\notE_T) & \sim &  M_h \, , \nonumber \\
M^*_T(c\ell\ell,\notE_T) & \sim &  m_t  \, ,
\end{eqnarray}
where $M^*$ is the value of invariant mass or cluster transverse mass 
with a peak of the distribution. 
These distributions provide powerful selection tools to remove physics 
background while maintaining the Higgs signal.

\section{Realistic Acceptance Cuts}

To study the discovery potential of this charming FCNH signal from top decays 
at the LHC, we have applied realistic basic cuts listed in Eq.~[22] 
and tagging efficiencies for $b-$jets.
In addition to basic cuts we apply cuts on invariant mass of jets 
and cluster transverse mass of $\ell\ell$ and $c\ell\ell$ 
to effectively veto the background events:
\begin{itemize}
\item[(a)] $|M_{jj} - m_W| \leq 0.15 \times m_W$,
\item[(b)] $|M_{bjj} - m_t| \leq 0.20 \times m_t$, 
\item[(c)] 50 GeV $\leq M_{T}(\ell\ell,\notE_T) \leq 150$ GeV, and 
\item[(d)] 100 GeV $\leq M_{T}(c\ell\ell,\notE_T) \leq 210$ GeV. 
\end{itemize}
These selection requirements remove more than 90\% of the total background. 


\begin{figure}[htb]
\begin{center}
\includegraphics[width=68mm]{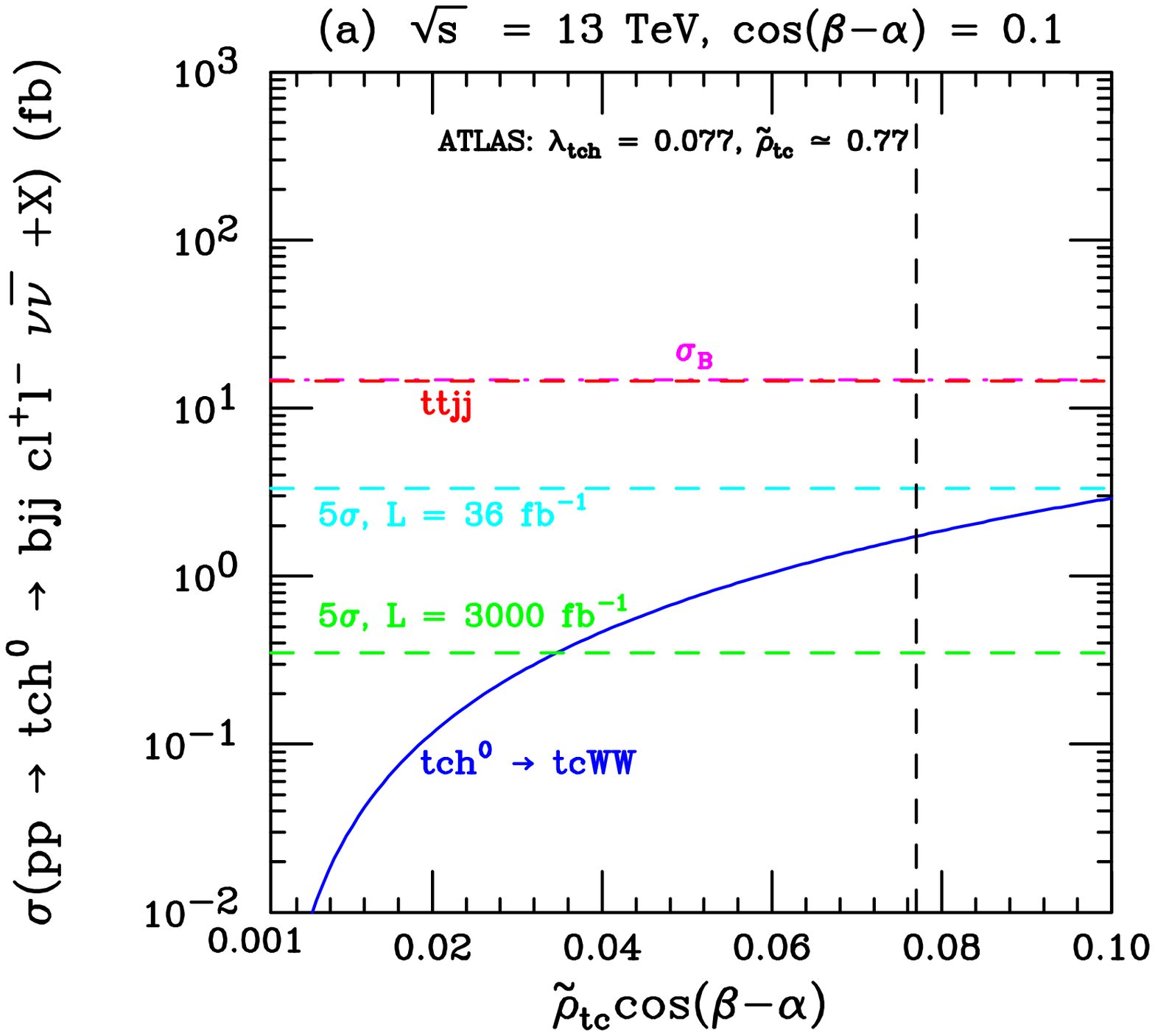}
\hspace{0.1in}
\includegraphics[width=68mm]{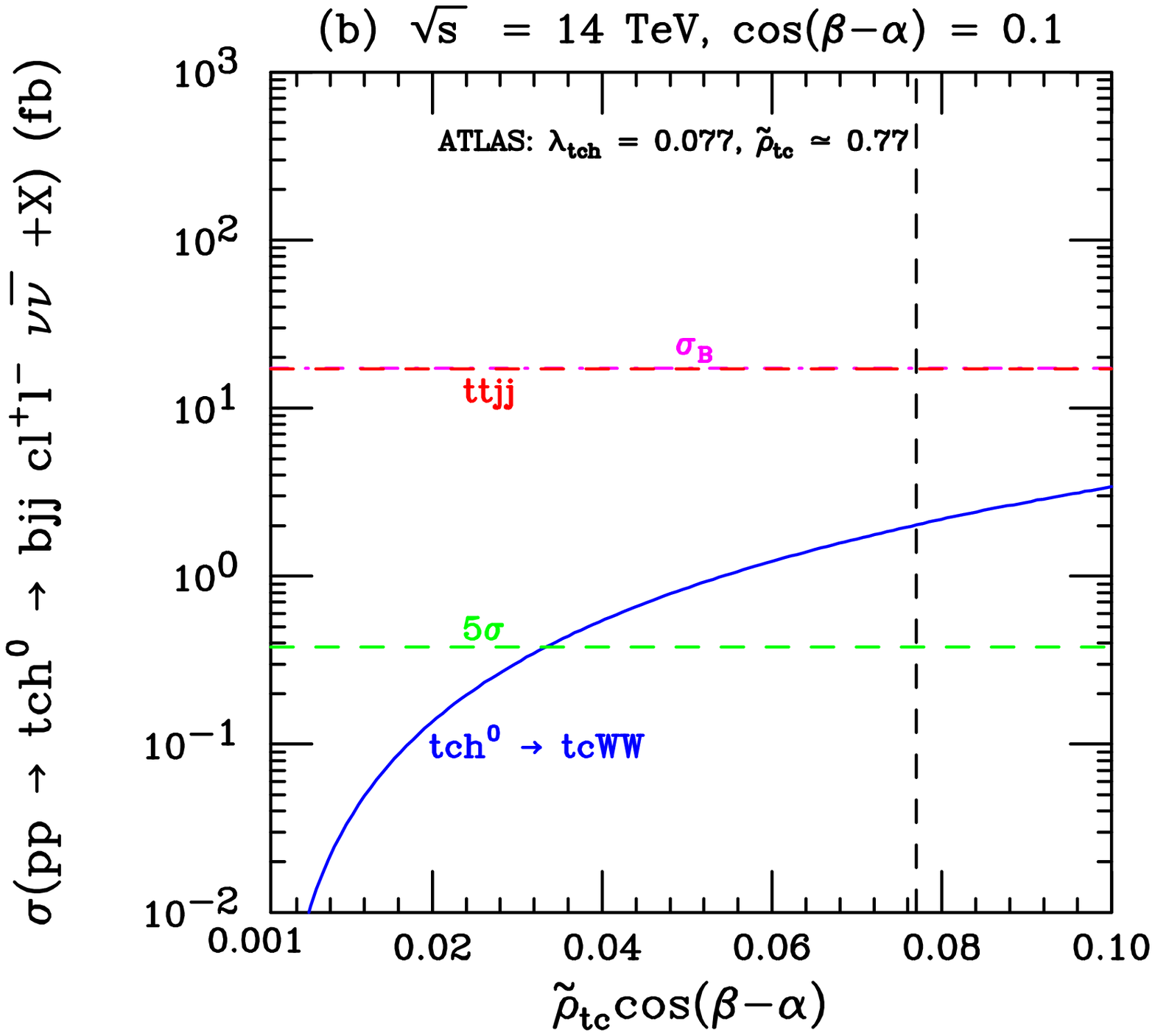}
\caption{
The cross section in fb of 
$pp \to t\bar{t} \to t c h^0 \to b j j c \ell^+ \ell^- +\notE_T +X$
at $\sqrt{s} =$ 13 TeV and 14 TeV as a function of $\tilde{\rho}_{tc}$, 
along with total (magenta dotdash) and most dominant (red dash) 
background after applying all the cuts, 
tagging and mistagging efficiencies and higher order QCD corrections.
The blue dash line and green dash line shows the minimum cross section 
needed for $5\sigma$ significance at $L = 36$ fb$^{-1}$ 
and 3 ab$^{-1}$ respectively for center of mass energy of 13 TeV. 
Where as for 14 TeV, we present $L = 3$ ab$^{-1}$ (green dash) only. 
The current ATLAS-Limit~\cite{Aaboud:2018pob} is shown as a black dash 
vertical line.
}
\label{fig:sigma13} 
\end{center}
\end{figure}

Measurement uncertainties in jet and lepton momenta as well as missing
transverse momentum give rise to a spread in the reconstructed masses
about the true values of $m_t$ and $M_\phi$.
Based on the ATLAS~\cite{ATLAS:2013-004} and the CMS~\cite{CMS:2016-0366} 
specifications we model these effects by Gaussian smearing of momenta:
\begin{eqnarray}
\frac{\Delta E}{E} = \frac{0.60}{\sqrt{E({\rm GeV})}} \oplus 0.03,
\end{eqnarray}
for jets and
\begin{eqnarray}
\frac{\Delta E}{E} = \frac{0.25}{\sqrt{E({\rm GeV})}} \oplus 0.01,
\end{eqnarray}
for charged leptons with individual terms added in quadrature.


\begin{figure}[htb]
\begin{center}
\includegraphics[width=68mm]{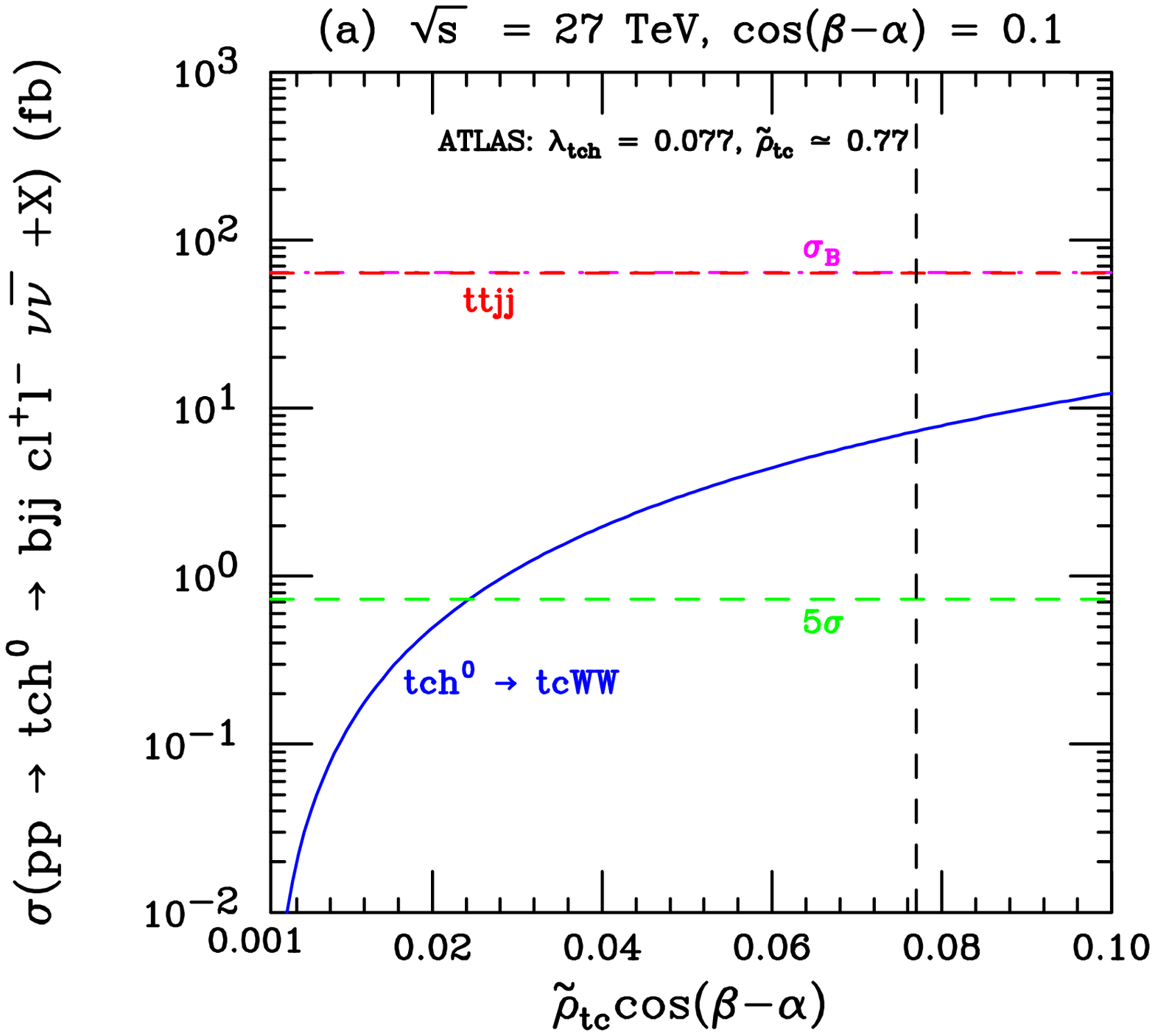}
\hspace{0.1in}
\includegraphics[width=68mm]{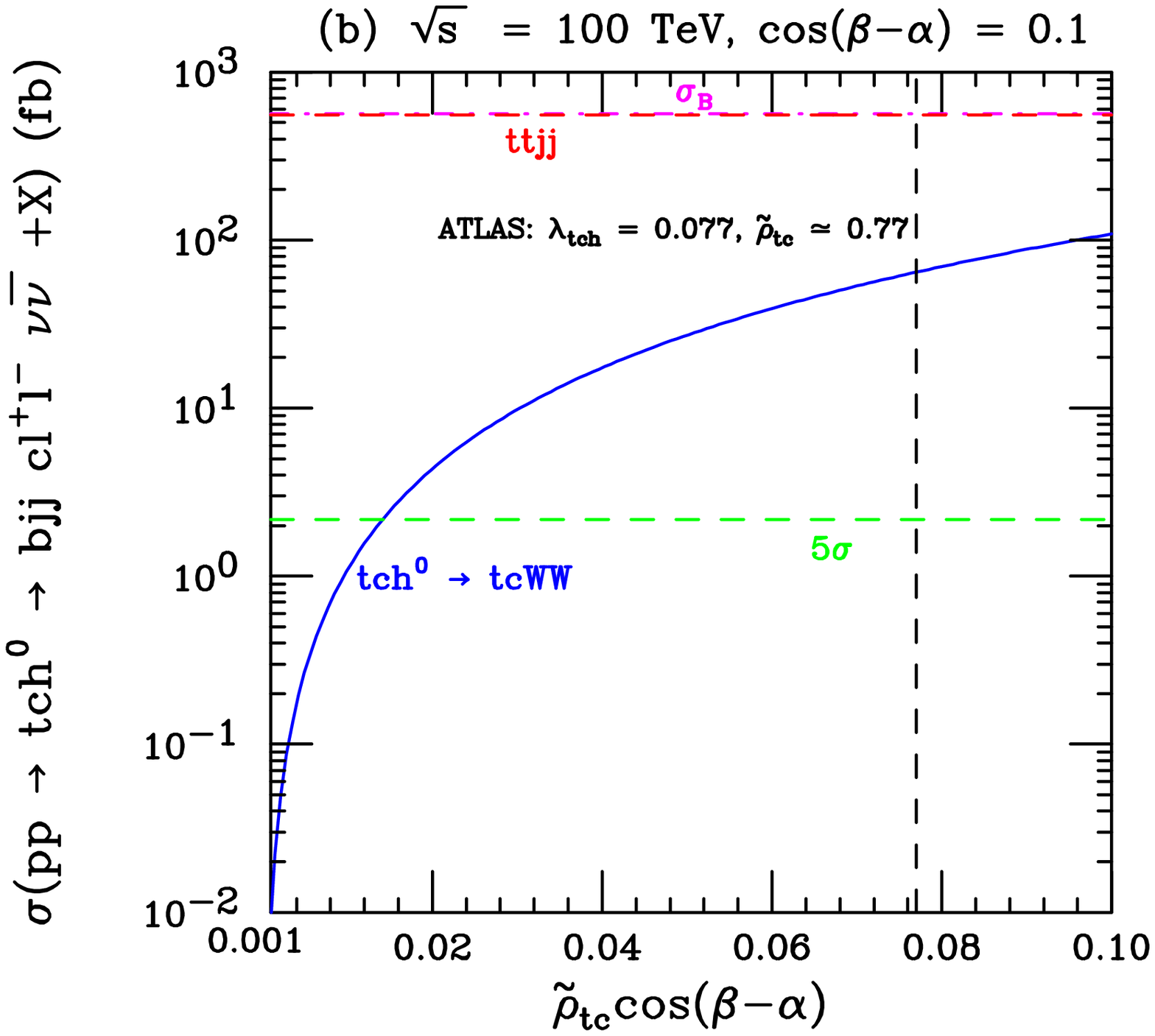}
\caption{Similar to FIG.~4, but for (a) $\sqrt{s} =$ 27 TeV, and (b) 100 TeV.
}

\label{fig:sigma27} 
\end{center}
\end{figure}

\section{Discovery Potential at the LHC}

Applying all realistic cuts, we present our results
for the Higgs signal at the LHC with 
$\sqrt{s} = 13$ TeV and $\sqrt{s} = 14$ TeV 
as well as 
cross sections for future hadron colliders with $\sqrt{s} = 27$ TeV and 
$\sqrt{s} = 100$ TeV in Table~II. 
Here we have kept $\cos(\beta - \alpha) = 0.1$. Later we will 
vary it from 0.01 to 0.2 for discovery contours.
Cross sections for dominant background processes are presented 
in Table~III.

\begin{table}[htb]
\label{sigma1tab}
\caption[]{Cross section of Higgs signal 
$pp \to t\bar{t} \to t c h^0 \to bjj \, c\ell\ell +\notE_T +X$ in fb with 
$\cos(\beta-\alpha) = 0.1$ for the LHC and future hadron colliders.}
\begin{tabular}{cllll}
\hline
\hline
$\tilde{\rho}_{tc}$ & 13 TeV  & 14 TeV  & 27 TeV  & 100 TeV \\
\hline
0.1  & $\;$ 0.015 & $\;$ 0.017 & $\;$ 0.06    & $\;$ 0.54  \\
\hline
0.5  & $\;$ 0.364  & $\;$ 0.425  & $\;$ 1.53     & $\;$ 13.6 \\
\hline
 1   & $\;$ 1.46  & $\;$ 1.70  & $\;$ 6.15     & $\;$ 54.4 \\
\hline
\hline
\end{tabular}
\end{table}
\begin{table}[htb]
\label{sigma2tab}

\caption[]{Cross section in fb for dominant physics background processes, 
with K factors and tagging efficiencies at the LHC and 
future hadron colliders.}
\begin{tabular}{cllll}
\hline\hline
Background & 13 TeV  & 14 TeV  & 27 TeV  & 100 TeV \\
\hline
ttjj   & $\;$ 14.6   & $\;$ 17.1    & $\;$ 63.6   & $\;$ 557 \\
\hline
ttW    & $\;$ 0.16   & $\;$ 0.17   & $\;$  0.36   & $\;$ 1.41 \\
\hline 
bbjj$\tau\tau$ & $\;$ 0.035   & $\;$ 0.039   & $\;$ 0.13& $\;$ 0.95 \\
\hline 
bbjjWW & $\;$ 0.003 & $\;$ 0.0035  & $\;$ 0.011  & $\;$ 0.09   \\
\hline  
ccjjWW & $\;$0.0017  & $\;$ 0.0019  & $\;$ 0.006  & $\;$ 0.05  \\
\hline
WWjjjj & $\;$ 9.96E-06 & $\;$ 1.12E-05   & $\;$ 2.48E-05    & $\;$ 0.0002 \\
\hline\hline
\end{tabular}
\end{table}

To estimate the discovery potential at the LHC we include curves that
correspond to the minimal cross section of signal ($\sigma_S$)
required by our discovery criterion described in the following.
We define the signal to be observable
if the lower limit on the signal plus background is larger than
the corresponding upper limit on the background 
with statistical fluctuations
\begin{eqnarray}
L (\sigma_S+\sigma_B) - N\sqrt{ L(\sigma_S+\sigma_B) } \ge
L \sigma_B +N \sqrt{ L\sigma_B },
\end{eqnarray}
or equivalently,
\begin{equation}
\sigma_S \ge \frac{N}{L}\left[N+2\sqrt{L\sigma_B}\right] \, .
\end{equation}
Here $L$ is the integrated luminosity,
$\sigma_S$ is the cross section of the FCNH signal,
and $\sigma_B$ is the background cross section.
The parameter $N$ specifies the level or probability of discovery.
We take $N = 2.5$, which corresponds to a 5$\sigma$ signal.

For $L\sigma_B \gg 1$, this requirement becomes similar to
\begin{eqnarray}
N_{\rm SS} = \frac{N_S}{\sqrt{N_B}}
 = \frac{L\sigma_S}{\sqrt{L\sigma_B}} \ge 5 \, ,
\end{eqnarray}
where
$N_S$ is the signal number of events,
$N_B$ is the background number of events,
and $N_{\rm SS}$ is the statistical significance, which is
commonly used in the literature.
If the background has fewer than 25 events for a given luminosity,
we employ the Poisson distribution and require that
the Poisson probability for the SM background to fluctuate to this
level is less than $2.87\times 10^{-7}$, i.e. an equivalent probability
to a 5-sigma fluctuation with Gaussian statistics.

Figure~4 shows the Higgs signal cross section as a function of 
$\tilde{\rho}_{tc}$, along with cross section of total background and 
the most dominant background process ($ttjj$) for the
CERN Large Hadron Collider with $\sqrt{s} = $13 and 14 TeV. 
We have also shown, minimum cross section required for 5$\sigma$ 
significance at $L = 36.1 fb^{-1}$ and higher luminosities for the future 
HL LHC~\cite{Barletta:2013ooa,Tomas:2016}, i.e 
$L =$ 300 and 3000 $fb^{-1}$. 
All tagging efficiencies and K factors discussed above are included. 
Our analysis suggests an improvement in the reach of ATLAS at a luminosity 
of 3000 $fb^{-1}$, which gets better at higher energies(HE-LHC), i.e 
$\sqrt{s} =$27 and 100 TeV, as shown in Figure ~5. 

We present the 5$\sigma$ discovery reach at the LHC for 
(a) $\sqrt{s} =$ 13 TeV and (b) $\sqrt{s} =$ 14 TeV in FIG.~6, 
in the parameter plane of $[\cos(\beta - \alpha),\tilde{\rho}_{tc}]$.
We have chosen $L =$ 300 and 3000 $fb^{-1}$.  
Figure~7 shows the discovery contours for $\sqrt{s} = $27 and 100 TeV. 
High energy (HE) LHC with high luminosity (HL) is quite promising 
as it nearly covers the entire parameter space that we have used 
in our analysis. 


\begin{figure}[htb]
\begin{center}
\includegraphics[width=68mm]{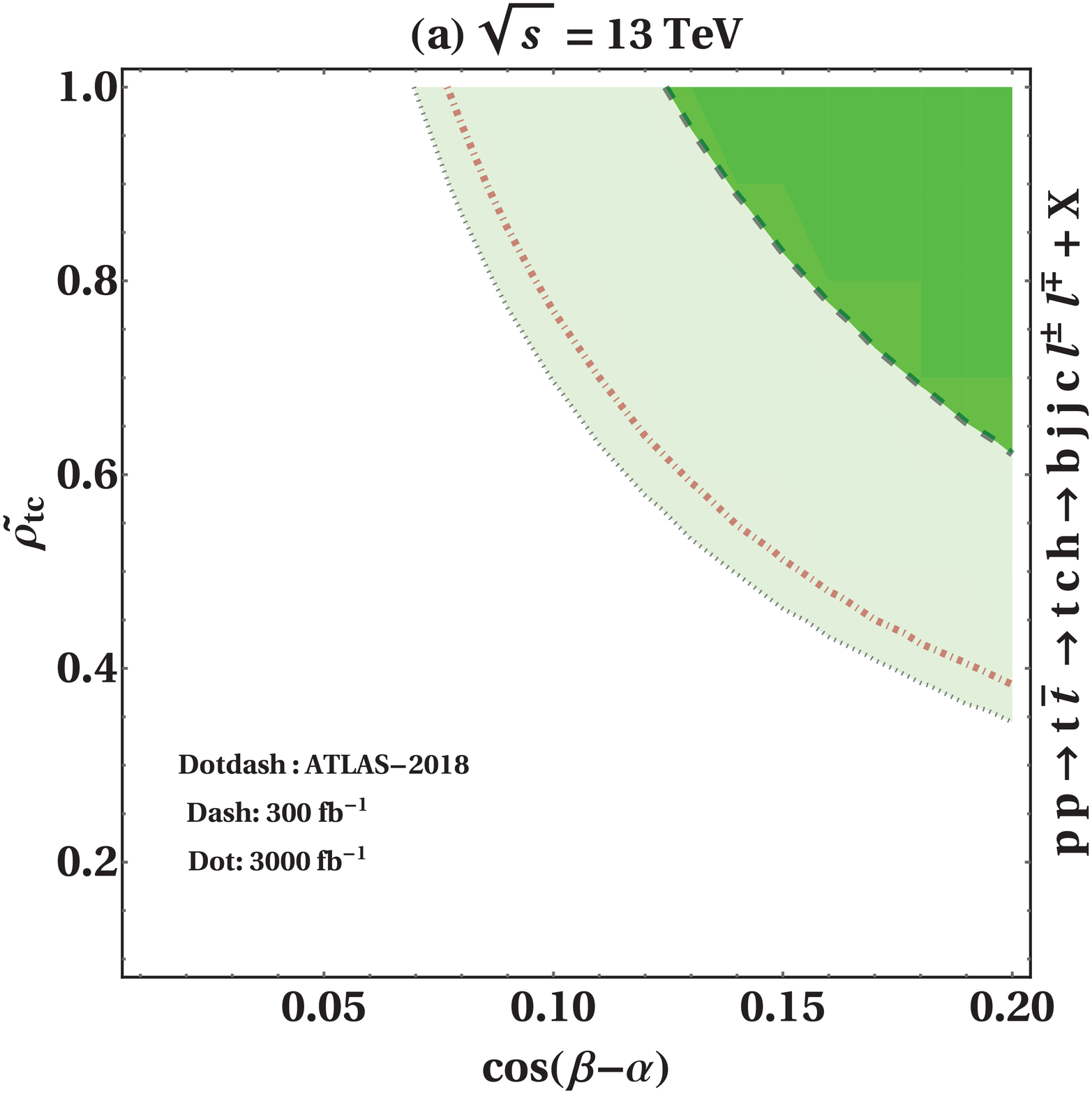}
\hspace{0.1in}
\includegraphics[width=68mm]{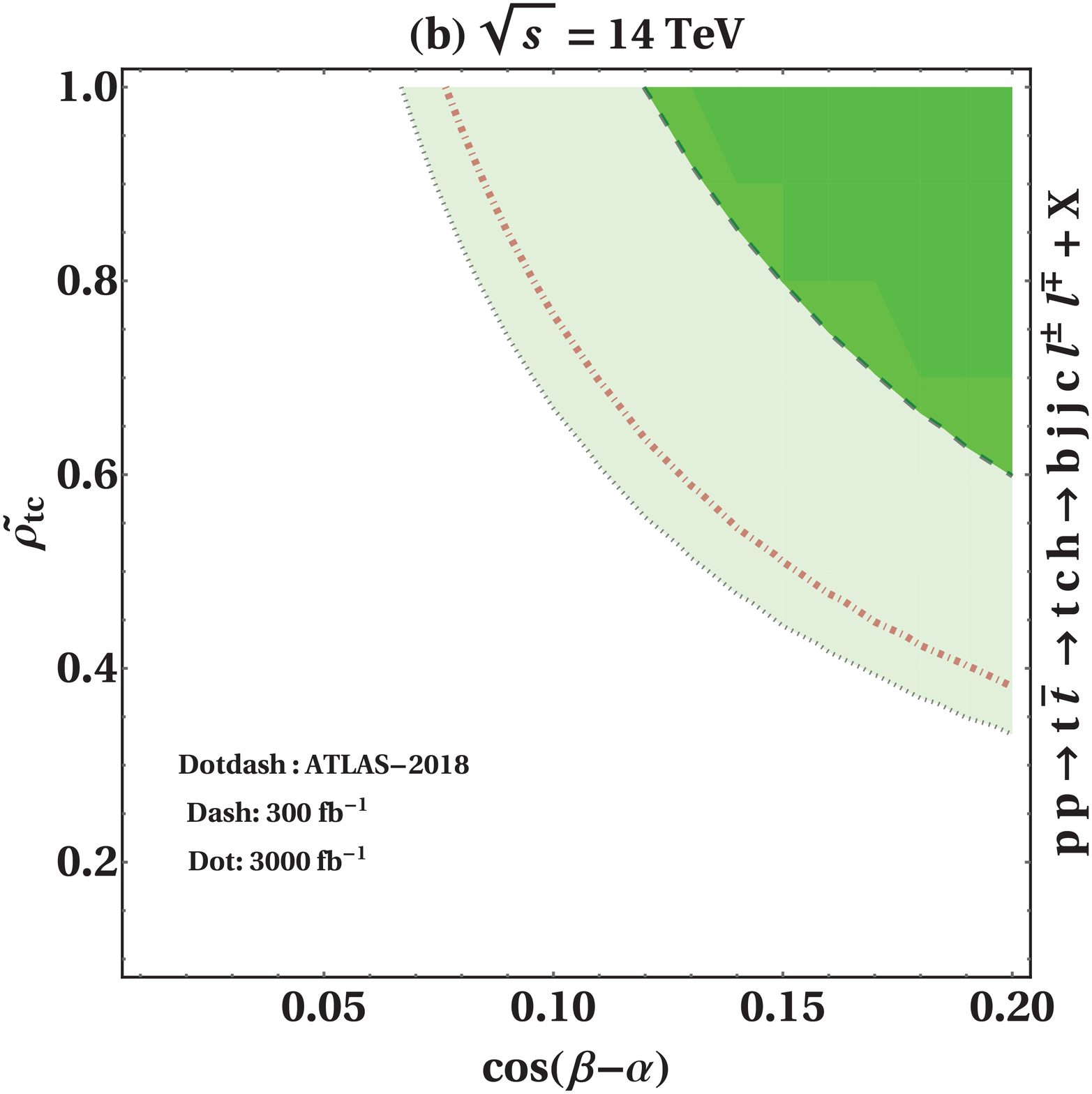}
\caption{
The 5$\sigma$ discovery contours at the LHC
in the plane of $[\cos(\beta - \alpha),\tilde{\rho}_{tc}]$
for (a) $\sqrt{s} =$ 13 TeV and (b) $\sqrt{s} =$ 14 TeV. 
For $L = 300 fb^{-1}$ (dash) and 
$L = 3000 fb^{-1}$ (dot).
Also shown is the current limit on 
$\lambda_{tch} = \tilde{\rho}_{tc}\cos(\beta-\alpha)$ (red dotdash) 
set by ATLAS~\cite{Aaboud:2018pob}.
The shaded region above this curve is excluded at 95\% CL.
}
\label{fig:contour1} 
\end{center}
\end{figure}


\begin{figure}[htb]
\begin{center}
\includegraphics[width=68mm]{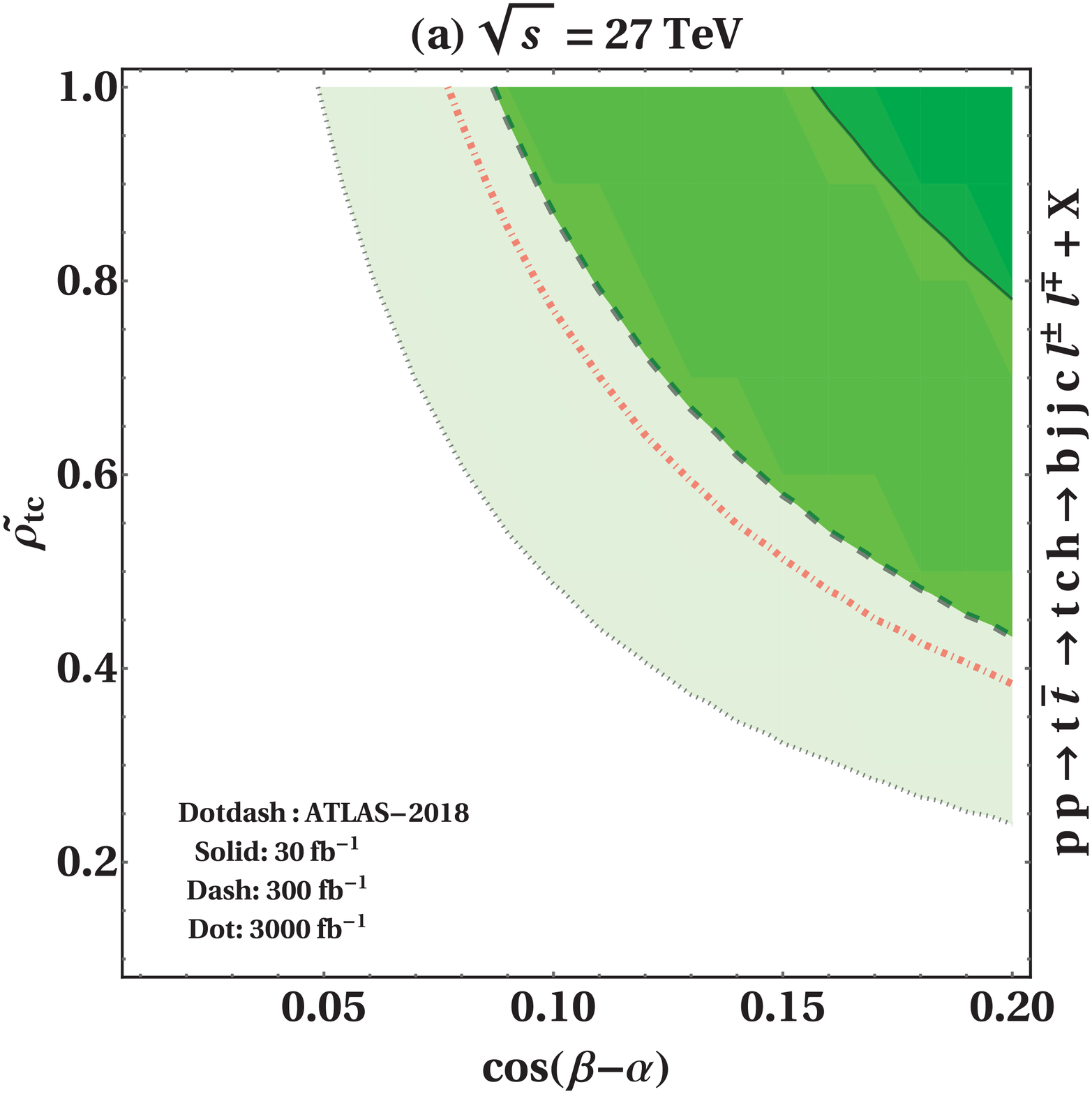}
\hspace{0.1in}
\includegraphics[width=68mm]{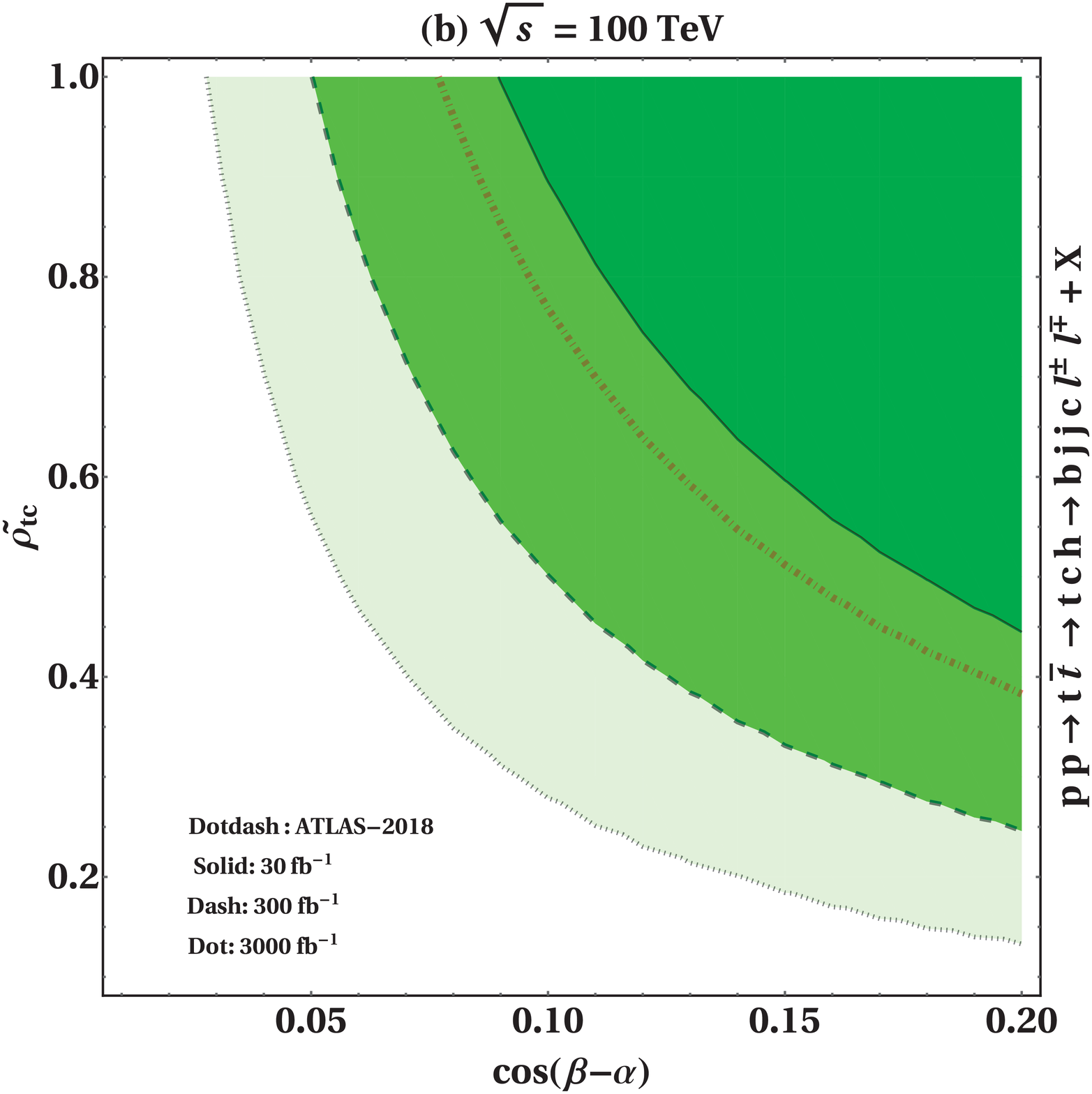}
\caption{The 5$\sigma$ discovery contours at future pp colliders 
in the plane of $[\cos(\beta - \alpha),\tilde{\rho}_{tc}]$
for (a) $\sqrt{s} =$ 27 TeV, and (b) $\sqrt{s} =$ 100 TeV, 
for $L = 30 fb^{-1}$ (solid), $L = 300 fb^{-1}$ (dash) 
and $L = 3000 fb^{-1}$ (dot).
Also shown is the current limit on 
$\lambda_{tch} = \tilde{\rho}_{tc}\cos(\beta-\alpha)$ (red dotdash) 
set by ATLAS~\cite{Aaboud:2018pob}.
The shaded region above this curve is excluded at 95\% CL.
}
\label{fig:contour2}  
\end{center}
\end{figure}

\section{Conclusions}

It is a generic possibility of particle theories beyond
the Standard Model to have contributions to tree-level FCNH interactions,
especially for the third generation quarks.
These contributions arise naturally in models with additional Higgs
doublets, such as the special two Higgs doublet model for the top quark 
(T2HDM), or a general 2HDM.
In the alignment limit, the light Higgs boson ($h^0$) resembles 
the standard Higgs boson, and it has a mass below the top mass. 
This could engender the rare decay $t \to c h^0$.

We investigated the prospects for discovering such a decay at the LHC,
focusing on the channel where $t\overline{t}$ are pair produced and
subsequently decay, one haronically and the other through the FCNH mode.
The primary background for this signal is a $t\overline{t} j j$ 
with both top quarks decaying leptonically.
This background involves one $b-$jet mis-tagged as a $c$ jet,and
 two other light jets, along with two leptons and missing transverse energy.
Nonetheless, by taking advantage of the available kinematic
information, we can reconstruct the resonances of the signal and
reject much of the background.

Based on our analysis, we find that LHC at $\sqrt{s} = 14$ TeV, 
with $L = 3000$ fb$^{-1}$, can probe to as low as 
${\cal B} (t \to c h^0) \simeq 1.17 \times 10^{-3}$ , 
$\lambda_{tch} = \tilde{\rho}_{tc}\cos(\beta-\alpha) \simeq 0.069$.
It gets better  with $\sqrt{s} = 27$ TeV and $\sqrt{s} = 100$ TeV, 
which can reach upto  ${\cal B}(t \to c h^0) \simeq 6.1 \times 10^{-4}$ , 
$\lambda_{tch} \simeq 0.048$ and  
${\cal B}(t \to c h^0) \simeq 2 \times 10^{-4}$ , 
$\lambda_{tch} \simeq 0.028$ respectively.

We look forward to being guided by more new experimental
results as we explore interesting physics of electroweak symmetry breaking 
(EWSB) and FCNH interactions.
While the properties of the Higgs boson goes under further scrutiny as 
data accumulate, perhaps a dedicated FCNH $t\to c h^0$ search 
should be undertaken, 
for upcoming HL LHC and further HE-LHC as well as future high energy 
hadron collider with a CM energy of 100 TeV.

\section*{Acknowledgments}

We are grateful to Kai-Feng Jack Chen for beneficial discussions.
C.K. thanks the Institute of Physics at the Academia Sinica and the 
High Energy Physics Group at National Taiwan University 
for excellent hospitality, where part of the research was completed.
This research was supported in part by the U.S. Department of Energy.


\end{document}